\documentclass[b5paper,12pt]{article}

\usepackage[margin=1.2cm]{geometry}
\usepackage{setspace}
\usepackage[affil-it]{authblk}
\usepackage{amsmath}
\usepackage{amssymb}
\usepackage{amsthm}
\usepackage{slashed}
\usepackage{psfrag}
\usepackage{graphicx,subfigure}
\usepackage{color}

\def\realeven{{\textrm{Re}\,\,\Lambda^*_+}}
\def\realodd{{\textrm{Re}\,\,\Lambda^*_-}}
\def\Realfull{{\textrm{Re}\,\,L^*}}
\def\Realeven{{\textrm{Re}\,\,L^*_+}}
\def\Realodd{{\textrm{Re}\,\,L^*_-}}
\def\Lmn{{\Lambda^\mu_{\,\,\,\,\nu}}}

\begin{document}
\title{A Modification on the Ivanenko-Landau-K\"ahler Equation}
\author{Min Gi Lee \thanks{\texttt{mgleemail@gmail.com} \\{\it Submitted to International Journal of Theoretical Physics}\\This research was supported by the project of National Junior research fellowship of the National Research Foundation of Korea under grant number 2011-0013447. }}
\affil{Department of Mathematical Science, KAIST, Daejeon, Korea}
\date{(Dated: \today)}
\maketitle
\begin{abstract}
  In this paper, we illustrated one scenario to modify the Ivanenko-Landau-K\"ahler equation. Since Ivanenko and Landau introduced the equation in 1928, the equation has been regarded as having a certain role as a fermion in particular in the discrete Lattice. Also, although it correctly is formulated as an alternative classical field equation by the Ideal projection for the Dirac equation in the Minkowski space-time, so does it only in that flat geometry. I. M. Benn and R. W. Tucker in 1985 and Yu. N. Obukhov and S. N. Solodukhin in 1994 suggested two resolutions respectively. They modified the equation in order for it to make senses as an alternative for the Dirac equation in the general space-time. This paper advances a still another approach, however in the Minkowski space yet, as the first stage toward the generalization. Two ingredients for the modifications are essential. One is the restriction of the space of anti-symmetric tensor field to only its subalgebras, not to its Ideals. The other is the modification on the mass term for the equation to mean the eigenvalue problem generating physical states. The Vector U(1) and the Axial U(1) symmetries built in the modified equation are described.
\end{abstract}
\section{Introduction and Preliminaries} \label{sec:intro}
We are concerned with the Ivanenko-Landau-K\"ahler equation and illustrate one possible modification to the equation. The Ivanenko-Landau-K\"ahler equation, ILK equation for brevity, is a relativistic field equation on anti-symmetric tensor fields, the inhomogeneous differential forms. The equation is introduced by Ivanenko and Landau \cite{ivanenko_d._zur_1928} in 1928 to describe the half-spin particles and re-discovered by K\"ahler in 1962 in \cite{kahler_innere_1962,kahler_innere_2011}. This equation has been inquired as an alternative of the Dirac equation as well as being another classical field equation for a few reasons. The operator is local that is to say the equation is a partial differential equation, and is of first order. Also it is Lorentz invariant and squares to the d'alembert $\Box$ operator. Together with the fact that the anti-symmetric tensor fields are geometrically obvious objects, the equation has received attention by many scholars. To borrow a phrase of Obukhov and Solodukhin, the research has a dramatic history, see Graf \cite{graf_differential_1978},
Benn and Tuchker \cite{benn_fermions_1983,benn_dirac_1985}, Budinich and Bugajska \cite{budinich_spinors_1985}, Rabin \cite{rabin_homology_1982}, Bullinaria \cite{bullinaria_kahler_1986}, Ivanenko and Obukhov \cite{ivanenko_gravitational_1985}, Obukhov and Solodukhin \cite{obukhov_reduction_1993,obukhov_dirac_1994}, Manko{\v{c}} Bor{\v{s}}tnik and Nielsen \cite{mankoc_unification_2000}. In particular, substantial study on the connection to the Lattice fermion has been investigated by many scholars, see Becher and Joos \cite{becher_dirac-kahler_1982}, Rabin \cite{rabin_homology_1982}, Mitra \cite{mitra_geometry_1983}, Bullinaria \cite{bullinaria_arbitrary_1985}, G\"ockeler and Joos \cite{gockeler_kahlers_1984}, Smalley \cite{smalley_1986}.


The core of all discussions on the ILK equation is best expressed in following quotation:
\begin{quote}
In Minkowski space K\"ahler's equation decouples into four minimal left Ideals of
the Clifford algebra, and is equivalent to four identical Dirac equations.
Thus the K\"ahler equation is not the Dirac equation. Moreover, in an arbitrary
space-time the K\"ahler equation does not split into minimal left Ideals. These
features of the K\"ahler equation are not inevitable consequences of using
differential forms for the description of half-integer spin, ...

\hfill(Benn and Tucker \cite{benn_dirac_1985})
\end{quote}
From this point of view the study to have a consistent field theory of half-spin particles with the approach in the arbitrary space-time could be reduced to the question whether one can project out the one corresponding part of the Dirac equation, for instance the one minimal left Ideal, both before and after the differential operator acts on. Also, this has to be done in global domain. Benn and Tucker \cite{benn_dirac_1985} and Obukhov and Solodukhin \cite{obukhov_dirac_1994} advanced two different resolutions to this problem.
Obukhov et al. \cite{obukhov_dirac_1994} restricted the fields in a subspace of the space of differential forms tangent to a certain 2-dimensional surface in 4-manifold.
In Benn and Tucker \cite{benn_dirac_1985}, they devised a spin-invariant inner product and a consequent action in each Ideal separately.

In this paper, we are going to advance our own modification. It will be a limited discussion in Minkowski space, however, for now and we are just in stage to suggest a new approach in this paper.

For the latter paper \cite{benn_dirac_1985}, there are two more things important for our study. One is the fact that they considered a combination of differential forms over $\mathbb{R}$. They pointed out that one does not have to complexify the fields to build a U(1) gauge theory. The other is the fact that the  algebra, the space of differential forms, is explicitly separable to its even and odd subalgebras, which we will specify in detail later.

This paper is organized as follows. Section \ref{sec:ILK} contains a brief review on the ILK theory. Section \ref{sec:even} provides grounds for our approach in intuitive level, and then Section \ref{sec:dke} specifies the modified action and the modified equation. In Section \ref{sec:spinor} we make a comparison our theory to the $4\times4$ matrix algebraic Spinors theory. In Section \ref{sec:cov} we put the most essential arguments on why the modified equation describes a half-spin particle and how the modified equation is covariant. Last section describes how the chirality is realized in the modified equation.
\section{The Ivanenko-Landau-K\"ahler Equation} \label{sec:ILK}
First we address that contents of this Section are largely borrowed from the review Section of Obukhov and Solodukhin \cite{obukhov_dirac_1994}.
Let $M^4$ be the Minkowski space-time with the metric $\eta_{\mu\nu} = \textrm{diag}(+1,-1,-1,-1)$. $\Lambda^*(M^4) = \bigoplus_{p=0}^{4} \Lambda^p(M^4)$ is a space of differential forms. $\Lambda^*(M^4)$ equipped with the wedge product $\wedge$, is referred to as the Exterior Algebra. Any element of the algebra is expressed as
\begin{equation}
\varphi = \sum_{p=0}^4 \frac{1}{p!}\varphi_{\mu_1\cdots\mu_p}\,\,dx^{\mu_1} \wedge \cdots \wedge dx^{\mu_p}, \label{eqn:genform}
\end{equation}
where $\{dx^\mu\}$ is the basis covector of cotangent space $T^*(M^4)$, dual to the coordinate basis $\{\partial_\mu\}$.

On $\Lambda^*(M^4)$, one introduces another product $\vee$, the Clifford product, by following formula introduced by K\"ahler \cite{kahler_innere_1962},
$$ dx^\mu \vee dx^\nu = dx^\mu \wedge dx^\nu + \eta^{\mu\nu}.$$
Then, under the assumption that there are no repetitions in indices $\mu_1,\cdots,\mu_p$, \eqref{eqn:genform} will be equivalently written as
\begin{equation}
\varphi = \sum_{p=0}^4 \frac{1}{p!}\varphi_{\mu_1\cdots\mu_p}\,\,dx^{\mu_1} \vee \cdots \vee dx^{\mu_p}. \label{eqn:genclif}
\end{equation}
$\Lambda^*(M^4)$ equipped with the product $\vee$ only, is called the Clifford algebra and with both of $\wedge$ and $\vee$, is called the K\"ahler-Atiyah algebra. The embedding between algebras associated is described in detail in Graf \cite{graf_differential_1978}.

$\Lambda^*(M^4)$ admits a differential operator $d$. For any $p$-form
\begin{align*}
  \omega &= \frac{1}{p!}\omega_{\mu_1\cdots\mu_p}\,\,dx^{\mu_1} \wedge \cdots \wedge dx^{\mu_p}, \\
  d \omega &=\frac{1}{p!}\partial_{\mu}\omega_{\mu_1\cdots\mu_p} \,\,dx^{\mu} \wedge dx^{\mu_1}\cdots \wedge dx^{\mu_p}.
\end{align*}

Also, with aid of the metric, $\Lambda^*(M^4)$ admits a Hodge-star operator $*$. We shall define the Hodge-star operator by
\begin{align}
  &* : \Lambda^p(M^4) \mapsto \Lambda^{4-p}(M^4), \quad \text{such that for any $p$-form $\omega$ and $\theta$,} \nonumber\\
  &*\omega \wedge \theta = dt\wedge dx\wedge dy\wedge dz\, (\omega,\theta). \label{eqn:*}
\end{align}
The pairing $(\omega,\theta)$ is defined as follows. If
\begin{align*}
  &\omega = \alpha_1 \wedge \cdots \wedge \alpha_p, \quad \theta = \beta_1 \wedge \cdots \wedge \beta_p, \quad\text{where $\alpha_i$ and $\beta_j$ are $1$-forms,}\\
  &(\omega,\theta) = \textrm{det}\,\,(\alpha_i,\beta_j), \quad\quad(\alpha_i,\beta_j):= (\alpha_i)_\mu \eta^{\mu\nu}(\beta_j)_\nu.
\end{align*}
Conventionally, the $*$ operator is defined by the formula
\begin{align*}
  \omega \wedge *\theta = dt\wedge dx\wedge dy\wedge dz\, (\omega,\theta)
\end{align*}
with $*$ on the right, but we adopt \eqref{eqn:*} in this paper. This can be understood as much as physicists put complex conjugated variable on the left in the inner product. The difference from the conventional one and ours are only in signs on odd order forms.

Now one introduces another differential operator $\delta$ on $\Lambda^*(M^4)$. There is a little deviation on the definition of $\delta$ from author to author, but we fixed our convention here as
$$\delta :=(-1)^p\,*^{-1}\,d\,*,$$
where $p$ is the order of differential form the operator acts on. 
We also address that the operator defined with the conventional Hodge star operator becomes identical to above. In this convention, the following are confirmed.
\begin{align*}
  &d^2 = \delta^2 = 0, \\
  &(d-\delta) = \partial_\mu \, dx^\mu \,\vee, \\
  &(d-\delta)^2 = -(d\delta + \delta d) = \partial_t^2 - \Delta,
\end{align*}
where $\Delta = \partial_x^2 + \partial_y^2 + \partial_z^2$, the Laplace operator.

In 1928, Ivanenko and Landau \cite{ivanenko_d._zur_1928} considered an equation for a field defined by \eqref{eqn:genform},
\begin{align}
  \big(i(d-\delta) - m\big) \varphi = 0. \label{eqn:ILK}
\end{align}
to describe the half-spin particles.
The equation is re-discovered by K\"ahler in 1962 \cite{kahler_innere_1962} and is often referred to as Dirac-K\"ahler equation. It should be mentioned that the definition of the equation varies among the scholars. Obukhov and Solodukhin \cite{obukhov_dirac_1994} used \eqref{eqn:ILK} as a starting point of their discussion but in many other literature, for example Becher and Joos \cite{becher_dirac-kahler_1982} or Benn and Tucker \cite{benn_fermions_1983,benn_dirac_1985} used
\begin{align}
  \big(d-\delta+ m\big) \varphi = 0. \label{eqn:DKE}
\end{align}
We also point out that the space $\varphi$ belongs to also varies from author to author. Most generally it is an element of $\Lambda^*(M^4)$ over $\mathbb{C}$  but Benn and Tucker first used the space in \cite{benn_fermions_1983} but later used $\Lambda^*(M^4)$ over $\mathbb{R}$ in \cite{benn_dirac_1985}. As was mentioned earlier, Obukhov restricted $\varphi$ in a subspaces of $\Lambda^*(M^4)$ related to a certain $2$-dimensional surface.

If $\varphi$ is an element of $\Lambda^*(M^4)$ over $\mathbb{C}$, then $\varphi$ is decomposed into elements of four minimal left Ideals. Here the algebraic structure we concern is the Cilfford product, we regard $\Lambda^*(M^4)$ the Clifford algebra. Now one can define four primitive Idempotents $\{P^{(i)}\}$ such that
\begin{align*}
  P^{(i)} \vee P^{(j)} &= 0, \quad i\ne j,\\
  P^{(i)} \vee P^{(i)} &= P^{(i)}, \quad \text{(no sum),}\\
  \sum_i P^{(i)} &= 1.
\end{align*}
Then
\begin{align*}
  \varphi = \sum_i \varphi^{(i)}, \quad \varphi^{(i)}:=\varphi\vee P^{(i)}.
\end{align*}
The primitive idempotents are constant idempotents in Minkowski space-time but are not in general space-time. This aspect was addressed in Section \ref{sec:intro} by quoting Benn and Tucker \cite{benn_dirac_1985}. This is the starting point of the previous works \cite{benn_dirac_1985,obukhov_dirac_1994} as to modify the equation since the original ILK equation only splits Ideals by the constant primitives. Without modification, one cannot single out the one Ideal component of the equation and thus cannot describe a single half-spin particle in general space-time.
\section{A Real Even/Odd Subalgebra} \label{sec:even}
This section is devoted to demonstrations of two things. One is a point of view that considering a combination of differential forms, either of even order or of odd order only, with real coefficients is an intriguing idea. The other is to attribute the necessity of concept of the minimal left Ideals, the main non-trivial issue in studying ILK equation, to the existence of mass term in the equation. Honestly, these are nothing to do with physics but are purely mathematical considerations.

It is helpful to define a few notations before moving on to the main task. For a given manifold $\mathcal{M}$, I will denote $\Lambda^*$ the $\Lambda^*(\mathcal{M})$ over $\mathbb{C}$ for simplicity and $\textrm{Re}\,\,\Lambda^*$ the algebra over $\mathbb{R}$. Also $\Lambda^*_+$ refers to as the even subalgebra $\bigoplus_{p\,\, \text{even}} \Lambda^p(\mathcal{M})$, and $\Lambda^*_-$ refers to as the odd subalgebra $\bigoplus_{p\,\, \text{odd}} \Lambda^p(\mathcal{M})$. By $\textrm{Re}\,\,\Lambda^*_{\pm}$, I mean the even and the odd subalgebra over $\mathbb{R}$.

Now we consider the Cauchy-Riemann equations in $2$-dimensions
\begin{align}
   \big(\partial_x-i\partial_y) \psi = 0. \label{eqn:CR}
\end{align}
Although issues on the Lattice fields do not come within the scope of this paper, we address a few things to deliver an intuitive explanation. Suppose we try to perform a numerical computation of \eqref{eqn:CR} in a Lattice. The most naive method to do it would be as follows. One complex number is assigned on each vertex of Lattice. Next, for approximation of \eqref{eqn:CR}, we compute the differences
$$ \frac{\psi(x+a,y)-\psi(x-a,y)}{2a} - i\frac{\psi(x,y+a)-\psi(x,y-a)}{2a} = 0,$$
centered at a certain vertex $(x,y)$, where $a$ is the spacing of the Lattice. We do not go further than telling this approximation significantly fails. This computation yields two independently pertaining lattice fields, which is the notorious doubling phenomenon in Lattice theory.

As an alternative, let us decouple the inquiry on $\psi$ to its real and imaginary part. If $\psi = u + i\,v$, $u$ and $v$ each a real-valued function, the Cauchy-Riemann equations read
\begin{align}
  u_x + v_y &=0, \label{eqn:div}\\
  -u_y + v_x &=0. \label{eqn:curl}
\end{align}

Now we return to differential forms. Let
\begin{align*}
  \varphi^{(e)} &:= -u + v\,dx\wedge dy, \\
  \varphi^{(o)} &:= u\,dx + v\,dy.
\end{align*}
The meaning of superscripts are clear, $\varphi^{(e)} \in \textrm{Re}\,\,\Lambda^*_+$, and $\varphi^{(o)} \in \textrm{Re}\,\,\Lambda^*_-$. If we operate $d-\delta$ to them,
\begin{align*}
  (d-\delta)\varphi^{(e)} = 0 &\Longleftrightarrow
  \begin{cases}
    \big(-\partial_x u - \partial_y v\big)dx = 0, \\
    \big(-\partial_y u + \partial_x v\big)dy = 0,
  \end{cases} \quad \text{and}\\
  (d-\delta)\varphi^{(o)} = 0 &\Longleftrightarrow
  \begin{cases}
    \big(\partial_x u  + \partial_y v\big) = 0, \\
    \big(-\partial_y u + \partial_x v\big)dx \wedge dy = 0.
  \end{cases}
\end{align*}
It is clear from above that the separation of the algebra to its even and odd subalgebra provides us a way to single out one set of Cauchy-Riemann equations.
\setcounter{subfigure}{0}
\begin{figure}[ht]
  \centering
  \subfigure[even discretization] {
     \psfrag{u1}{$v_1$}  \psfrag{u2}{$v_2$}  \psfrag{u3}{$v_3$}  \psfrag{u4}{$v_4$}  \psfrag{u5}{$v_5$}
     \psfrag{u6}{$v_6$}  \psfrag{u7}{$v_7$}  \psfrag{u8}{$v_8$}  \psfrag{u9}{$v_9$}
     \psfrag{v1}{$-u_1$}  \psfrag{v2}{$-u_2$}  \psfrag{v3}{$-u_3$}  \psfrag{v4}{$-u_4$}  \psfrag{v5}{$-u_5$}
     \psfrag{v6}{$-u_6$}  \psfrag{v7}{$-u_7$}  \psfrag{v8}{$-u_8$}  \psfrag{v9}{$-u_9$}
     \includegraphics[width=4cm]{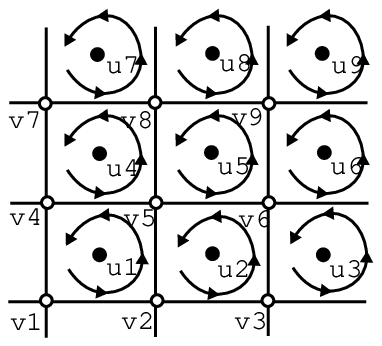} \label{fig:1a}
  }
  \quad\quad\quad\quad
  \subfigure[odd discretization] {
     \psfrag{u1}{$u_1$}  \psfrag{u2}{$u_2$}  \psfrag{u3}{$u_3$}  \psfrag{u4}{$u_4$}  \psfrag{u5}{$u_5$}
     \psfrag{u6}{$u_6$}  \psfrag{u7}{$u_7$}  \psfrag{u8}{$u_8$}  \psfrag{u9}{$u_9$}
     \psfrag{v1}{$v_1$}  \psfrag{v2}{$v_2$}  \psfrag{v3}{$v_3$}  \psfrag{v4}{$v_4$}  \psfrag{v5}{$v_5$}
     \psfrag{v6}{$v_6$}  \psfrag{v7}{$v_7$}  \psfrag{v8}{$v_8$}  \psfrag{v9}{$v_9$}
     \includegraphics[width=4cm]{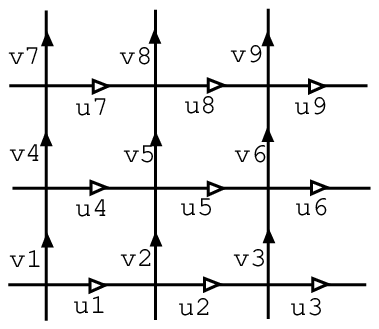} \label{fig:1b}
  }
  \caption{Two discretizations of the Cauchy-Riemann equation}
\end{figure}
In Lattice, one can perform the approximation keeping the geometric meaning by replacing each $p$-form by $p$-th order cochain in Lattice. This subject is introduced to this area by Becher and Joos \cite{becher_dirac-kahler_1982} and Rabin \cite{rabin_homology_1982}. In this discrete languages, the values of $p$-th order cochains are assigned on $p$-th order chains. $0$-th order chain refers to as vertices, $1$st order chain refers to as edges and $2$nd order chain refers to as the small plaquettes or faces and so on. The terminology such as chain, cochain are of Algebraic Topology and we will not discuss them in detail. We address that the numerical schemes this consideration provides reflect harmonic and balancing properties between involved quantities.  For example, it is well-known the case in figure \ref{fig:1b} describes the balance of divergence and the balance of curl of the vector field $(u(x),v(x))$.
The divergence equation comes from inspecting the net outgoing edge values on each vertex, and the curl equation comes from summing over the edge values along counter-clockwise direction on each loop or plaquette. The difference equation with indices in \ref{fig:1b}
\begin{align*}
\frac{1}{a}\big(u_5 - u_4 + v_5 - v_2\big) &= 0\quad\text{for approximation of divergence,}\\
\frac{1}{a}\big(u_1 + v_2 - u_4 - v_1\big) &= \frac{1}{a}\big(-(u_4-u_1) + v_2-v_1\big) =0\quad\text{for approximation of curl.}
\end{align*}
We can do similar with even order forms by inspecting differences of cochain fields with appropriate signs given by the designated arrows as in figure \ref{fig:1a}. The success of these numerical scheme in Lattice, together with the failure of the naive one earlier, supports our attempt to translate to the differential forms restricted in subalgebras.

Having got the $2$-dimensional case, which has nothing to do with physics, we may now turn to the $4$-dimensional Dirac equation. As was done in the Cauchy-Riemann equations, let us decouple inquiry on the Dirac Spinor to each real and imaginary part. Since it is $\mathbb{C}^4$ vector, it contains $8$ real-valued functions. The dimensions of real $p$-forms are $(1,4,6,4,1)$ in increasing order from $0$ to $4$. There are $8$ dimensions for the even real subalgebra and $8$ dimensions for the odd real subalgebra. It seems reasonable to suppose the only half of the algebra over $\mathbb{R}$ is responsible for the Dirac equation as was in Cauchy-Riemann equations. Indeed let us mention that the massless Dirac equation, $i\partial_\mu \gamma^\mu \psi = 0$ translates to $(d-\delta)\varphi^{(e)}=0$, or $(d-\delta)\varphi^{(o)}=0$ by the very way we did on Cauchy-Riemann equations. The full description is taken up in the Section \ref{sec:spinor}. The Cauchy-Riemann equations and the massless Dirac equation are differ only in dimensions and number of degrees of freedom in this aspects. Numerical implementation on Lattice for the massless Dirac equation works perfectly but this inquiry lies outside the scope of this paper.

To consider a real even or a real odd subalgebra is not only a personal presumption but reflecting the following particular backgrounds. It is well-known that the Dirac Spinor represents the Lorentz group. In particular, it represents the group $Spin(1,3)$ which doubly covers the restricted Lorentz group, $SO^{\uparrow}(1,3)$. It is also well-known that the group $Spin(1,3)$ are represented by the real even Clifford subalgebra $C\ell_+(1,3)$ or equivalently by the odd real subalgebra $C\ell_-(1,3)$. These considerations support our presumption to build an equation with only real and half of differential forms.

Now we are in a position to examine the mass term. Let us report a simple observation that $\varphi^{(e)}$, an element of $\Lambda^*_+$, never be a solution of ILK equation. The operator $d$ increases one order of differential forms and $\delta$ decreases one order of differential forms, in other words $(d-\delta)$ turns $\varphi^{(e)}$ into odd forms. On the while the mass term $m\varphi^{(e)}$ remains to be even thus the two terms cannot be equated. This applies for $\varphi^{(o)}$ either. To put it the other way round, the ILK equation calls for a combination of both even and odd forms for its solution. If we do not restrict the space of differential forms to the half of the algebra, the algebra now has $4$ minimal left Ideals which correspond $4$ Dirac equations. We need the primitive Idempotents and recover the classical approach.

What I have tried to show is that the subalgebra separation together with considering only massless equation shows mathematically intriguing trivial structures. Beyond this subalgebra separation, the minimal left Ideal separation is necessary only if the mass term is introduced. One may notice that, however, the philosophy to keep one Dirac component in same mathematical substructure, the minimal left Ideal or the subalgebra, as an objective of modification is kept in the approach.
\section{A Modified Field Equation and An Action} \label{sec:dke}
Now we propose the modified equation in Minkowski space. In the equation, we have degrees of freedom. Let $\bar\theta$ be an arbitrary constant space-like $1$-form field with unit norm, to be precise, $(\bar\theta)_\mu \eta^{\mu\nu} (\bar\theta)_\nu = -1$. Consider the field equation
\begin{align}
  (d-\delta)\varphi +m\,\varphi\vee(*\bar\theta) = 0, \quad \text{(the modified equation 1),}\label{eqn:dke}
\end{align}
where $\varphi$ is restricted in $\realeven$.

Since $\bar\theta$ is space-like, there must be an observer whose coordinate basis $dz=\bar\theta$. The equation \eqref{eqn:dke} is set to be identical with the Dirac equation with Dirac base for this observer. We will verify this in the Section \ref{sec:spinor} and here we draw attention to the U(1) symmetry in this equation.
\subsection{U(1) Gauge theory}
In Benn and Tucker \cite{benn_dirac_1985}, they pointed out that the real algebra formulation also admits U(1) symmetry. They used both even and odd forms but over $\mathbb{R}$ and implemented the symmetry using the volume form $dV = dt\wedge dx\wedge dy \wedge dz$, with aid of the property $dV\vee dV=-1$. In our discussion, we use only elements of $\realeven$ or $\realodd$ and implement the symmetry in different way.

For the observer whose coordinate basis $dz=\bar\theta$,
$$*\bar\theta = *dz = -dt\vee dx\vee dy.$$ Define
\begin{align}
  &\chi=dy\vee dx \quad \text{and} \quad \chi'=\chi^{-1}=-\chi=dx \vee dy, \nonumber
  \intertext{then}
  &\chi\vee\chi = \chi'\vee\chi'=-1, \quad dt \vee \chi = \chi \vee dt = *dz. \label{eqn:tauchi}
\end{align}
Now if $\varphi$ is a solution of the \eqref{eqn:dke}, define $\tilde\varphi = \varphi \vee \chi$, then
\begin{align*}
  (d-\delta)\varphi + m\varphi\vee(*dz) = 0 &\Longleftrightarrow (d-\delta)(\tilde\varphi\vee\chi') + m\tilde\varphi \vee \chi' \vee *dz = 0, \\
  \times \chi&\Longleftrightarrow (d-\delta)\tilde\varphi + m\tilde\varphi \vee \chi' \vee *dz \vee \chi = 0, \\
  &\Longleftrightarrow (d-\delta)\tilde\varphi + m\tilde\varphi \vee \chi' \vee \chi \vee dt \vee \chi = 0, \\
  &\Longleftrightarrow (d-\delta)\tilde\varphi + m\tilde\varphi\vee*dz = 0.
\end{align*}
Therefore $\tilde\varphi = \varphi \vee \chi$ is a solution and also is a $\varphi\vee\big(\cos{\lambda}\,1 + \sin{\lambda}\,\chi)$ for any $\lambda\in\mathbb{R}$ thus we have U(1) symmetry.

Let us denote $dt$ for this observer by $\tau$ for a general observer and write $(*\bar\theta) = \tau \vee \chi$. Then \eqref{eqn:dke} is equivalent to
\begin{align}
  (d-\delta)\varphi \vee \chi - m \varphi \vee \tau &= 0, \nonumber \\
  \Longleftrightarrow\partial_\mu \, dx^\mu \,\vee \varphi \vee\chi - m \varphi \vee \tau &= 0. \label{eqn:dke2}
\end{align}
Here, without the attached $\tau$ in the mass term, the form of the equation is parallel to the Dirac equation, $dx^\mu$ can be compared to $\gamma^\mu$ and $\chi$ to $i$.

This equation is the Euler-Lagrange equation of the action
\begin{align}
  \int *\phi' \wedge \big((d-\delta)\varphi\vee\chi\big) - m*\phi' \wedge \varphi', \label{eqn:dkeaction}
\end{align}
where $(\,\cdot\,)'=(\,\cdot\,)\vee(\tau)$. We intentionally did not project the integrand to the $\Lambda^4(M^4)$, the densities, because integration naturally ignores forms of order less than $4$. We ignore the interactions on measure zero sets.

At the present stage, we introduce $A \in \Lambda^1(M^4)$ to the equation for an electro-magnetic interaction.
\begin{align}
  &\partial_\mu \, dx^\mu \,\vee \varphi \vee\chi -eA \vee \varphi - m\varphi \vee \tau = 0, \quad \text{(the modified equation 2).}\label{eqn:dkeEM}
\end{align}
As usual, we have a gauge freedom on $\delta A$.


Now it is easy to verify this equation admits the U(1)-Gauge symmetry,
\begin{align*}
  \varphi &\mapsto \varphi\vee\big(\cos{\lambda(x)}\,1 + \sin{\lambda(x)}\,\chi), \\
  A_\mu &\mapsto A_\mu + \partial_\mu \lambda(x).
\end{align*}

So far, we used $\varphi \in \realeven$ in this Section, but equation in $\realodd$ also is treated incidently. By Clifford multiplication of $\tau$ on the right of the equation \eqref{eqn:dkeEM}, and using \eqref{eqn:tauchi}, one obtains
\begin{align*}
  &\partial_\mu \, dx^\mu \,\vee (\varphi\vee\tau) \vee\chi -eA \vee (\varphi\vee\tau) - m(\varphi \vee \tau)\vee\tau = 0.
\end{align*}
Since $\varphi\vee\tau$ is an arbitrary element of $\realodd$, what we obtained is the same equation.

We are left to answer to the two questions. One is that why the modified equation describes a half-spin particle and how the modified equation is covariant with existence of the free parameter $\bar\theta$. In other words, why our equation is not {\it ad-hoc}. The other is about the danger
of replacing $i$ with $\chi$ because it alters the scalar $i$ to the $2$-form $\chi$, which may cause problems in general space-time. For the former question, we shall give the answer in Section \ref{sec:cov}. For the latter question, however, it is beyond the author's knowledge and will not be discussed. 
\section{Comparison to Algebraic Spinors} \label{sec:spinor}
In this section, we shall discuss about the precise connection between Spinor formulation and our anti-symmetric tensor field formulation. 
\subsection{$4\times4$ matrix Spinors and Dirac Base}
Let us mention that the materials here are again borrowed from Obukhov and Solodukhin \cite{obukhov_dirac_1994}. The algebraic Spinors written in $4\times4$ complex matrix were discussed in Ivanenko and Obukhov \cite{ivanenko_gravitational_1985}.
In parallel with \eqref{eqn:genform}, let us consider a $4\times4$ matrix
\begin{align}
  \Psi = \sum_{p=0}^4 \frac{1}{p!}\Psi_{\mu_1\cdots\mu_p}\,\,\Gamma^{\mu_1\cdots\mu_p}, \label{eqn:gen4x4}
\end{align}
where $\Gamma^{\mu_1\cdots\mu_p} = \gamma^{[\mu_1}\cdots\gamma^{\mu_p]}$. We are using the Dirac base. Under the assumption that there are no repetitions in the indices $\mu_1,\cdots,\mu_p$, $\frac{1}{p!} \Gamma^{\mu_1\cdots\mu_p} = \gamma^{\mu_1}\cdots\gamma^{\mu_p}$ and since we do not want to introduce another notations, let us assume that there are no repeated indices in the expression $\gamma^{\mu_1}\cdots\gamma^{\mu_p}$ through out in this paper and use only lower case character. These are the replacement of $dx^{\mu_1} \vee \cdots \vee dx^{\mu_p}$ bearing in mind that $\{\gamma^\mu\}$ and $\{dx^\mu\}$ are base of Clifford algebras formulated in different languages. Again, it is helpful to define notations for the matrix subalgebras parallel to the $\Lambda^*$, $\Lambda^*_+,\cdots$ of differential forms with $L^*$, $\Realfull$, $L^*_+$, $L^*_-$, $\Realeven$ and $\Realodd$. The definitions are clear.

It is important that $4\times4$ matrix Spinor $\Psi$ admits the tensor-like transformation. Under the Lorentz group transformations
\begin{align*}
  x'^{\mu} = \Lmn x^\nu
\end{align*}
and $S[\Lmn]$ the associated representation of $\Lmn$ acting on the column Spinors, we have a formula
\begin{align*}
  S[\Lmn]^{-1}\gamma^\mu S[\Lmn] = \Lmn\gamma^\nu.
\end{align*}
For $\Phi_{(p)} = \Psi_{\nu_1\cdots\nu_p}\,\,\gamma^{\nu_1\cdots\nu_p}$, 
\begin{align*}
  S\Phi_{(p)} S^{-1}
  &= \Psi_{\nu_1\cdots\nu_p}S\gamma^{\nu_1\cdots\nu_p}S^{-1} \\
  &= \Psi_{\nu_1\cdots\nu_p}S\gamma^{\nu_1}S^{-1}S\gamma^{\nu_2}S^{-1}\cdots S\gamma^{\nu_p}S^{-1}\\
  &= \Psi'_{\mu_1\cdots\mu_p} \Lambda^{\mu_1}_{\,\,\,\,\nu_1}\cdots\Lambda^{\mu_p}_{\,\,\,\,\nu_p}\,\,S\gamma^{\nu_1}S^{-1}S\gamma^{\nu_2}S^{-1}\cdots S\gamma^{\nu_p}S^{-1}\\
  &=\Psi'_{\mu_1\cdots\mu_p}\,\,\gamma^{\mu_1\cdots\mu_p},
\end{align*}
where $S = S[\Lmn]$.
Therefore we see $\Psi'_{\mu_1\cdots\mu_p}$, the coefficients of $\Psi'=S\Psi S^{-1}$, transforms as covariant tensor.

$\Psi$ is a $4\times4$ complex matrix and one considers one left minimal Ideal component by multiplying $f=\begin{pmatrix} 1\\0\\0\\0 \end{pmatrix}$ so that $\psi=\Psi\,f$ to become a column Spinor. This is a usual treatment as in Chevalley \cite{chevalley_algebraic_1997}. Note that, however, we are giving a geometric meaning on $\Psi f$ by set $\Psi$ as in its right-hand-side in \eqref{eqn:gen4x4}.

It is presumably used in the literature that the $L^*$ and the $\Lambda^*$ are almost equivalent. $\Psi \in L^*$ given by the formula \eqref{eqn:gen4x4} can be an arbitrary $4\times4$ complex matrix and that reversely, an arbitrary matrix has the unique decomposition to a form of the right-hand-side of \eqref{eqn:gen4x4}. As a consequence, the combinations \eqref{eqn:gen4x4} directly translates to the combinations in the $\Lambda^*$. We shall look more carefully into the subject because we want to utilize the intermediate ingredients later. To be concrete, we examine the base $\gamma^{\mu_1\cdots\mu_p}$.

First thing we observe is that either in $\Realeven$ or in $\Realodd$, each basis does not share any non-zero position at the matrix entries to each other. We count the real and imaginary part differently because we are considering real coefficients.
For examples, base in odd subalgebra are
\begin{alignat}{4} \label{eqn:oddbase}
  \gamma^0 &= \begin{pmatrix}1 & 0 & 0 & 0\\0 & 1 & 0 & 0\\0 & 0 & -1 & 0\\0 & 0 & 0 & -1\end{pmatrix},
  & \gamma^0\gamma^2\gamma^1 = \begin{pmatrix}i & 0 & 0 & 0\\0 & -i & 0 & 0\\0 & 0 & -i & 0\\0 & 0 & 0 & i\end{pmatrix}, \nonumber\\
  \gamma^0\gamma^3\gamma^1 &= \begin{pmatrix}0 & -1 & 0 & 0\\1 & 0 & 0 & 0\\0 & 0 & 0 & 1\\0 & 0 & -1 & 0\end{pmatrix},
  &\gamma^0\gamma^3\gamma^2 = \begin{pmatrix}0 & i & 0 & 0\\i & 0 & 0 & 0\\0 & 0 & 0 & -i\\0 & 0 & -i & 0\end{pmatrix},\nonumber\\
  -\gamma^3 &= \begin{pmatrix}0 & 0 & -1 & 0\\0 & 0 & 0 & 1\\1 & 0 & 0 & 0\\0 & -1 & 0 & 0\end{pmatrix},
  ~~~~&-\gamma^3\gamma^2\gamma^1 = \begin{pmatrix}0 & 0 & -i & 0\\0 & 0 & 0 & -i\\i & 0 & 0 & 0\\0 & i & 0 & 0\end{pmatrix},\nonumber\\
  -\gamma^1 &= \begin{pmatrix}0 & 0 & 0 & -1\\0 & 0 & -1 & 0\\0 & 1 & 0 & 0\\1 & 0 & 0 & 0\end{pmatrix},
  &-\gamma^2 = \begin{pmatrix}0 & 0 & 0 & i\\0 & 0 & -i & 0\\0 & -i & 0 & 0\\i & 0 & 0 & 0\end{pmatrix}.
\end{alignat}
We assigned signs on the base in order for their first columns to have positive signs.
In conclusion, under the assumption that $\Psi\in\Realodd$, terms with different base are not {\it like terms}. The same rule applies for $\Realeven$. The base in even subalgebra are
\begin{alignat}{4} \label{eqn:evenbase}
  \textrm{I} &= \begin{pmatrix}1 & 0 & 0 & 0\\0 & 1 & 0 & 0\\0 & 0 & 1 & 0\\0 & 0 & 0 & 1\end{pmatrix},
  &
  \gamma^2\gamma^1 &= \begin{pmatrix}i & 0 & 0 & 0\\0 & -i & 0 & 0\\0 & 0 & i & 0\\0 & 0 & 0 & -i\end{pmatrix}, \nonumber\\
  \gamma^3\gamma^1 &= \begin{pmatrix}0 & -1 & 0 & 0\\1 & 0 & 0 & 0\\0 & 0 & 0 & -1\\0 & 0 & 1 & 0\end{pmatrix},
  &
  \gamma^3\gamma^2 &= \begin{pmatrix}0 & i & 0 & 0\\i & 0 & 0 & 0\\0 & 0 & 0 & i\\0 & 0 & i & 0\end{pmatrix},\nonumber\\
  \gamma^0\gamma^3 &= \begin{pmatrix}0 & 0 & 1 & 0\\0 & 0 & 0 & -1\\1 & 0 & 0 & 0\\0 & -1 & 0 & 0\end{pmatrix},
  &
  ~~~~\gamma^0\gamma^3\gamma^2\gamma^1 &= \begin{pmatrix}0 & 0 & i & 0\\0 & 0 & 0 & i\\i & 0 & 0 & 0\\0 & i & 0 & 0\end{pmatrix},\nonumber\\
  \gamma^0\gamma^1 &= \begin{pmatrix}0 & 0 & 0 & 1\\0 & 0 & 1 & 0\\0 & 1 & 0 & 0\\1 & 0 & 0 & 0\end{pmatrix},
  &
  \gamma^0\gamma^2 &= \begin{pmatrix}0 & 0 & 0 & -i\\0 & 0 & i & 0\\0 & -i & 0 & 0\\i & 0 & 0 & 0\end{pmatrix}
\end{alignat}
As consequences, the general combination in the subalgebras with real coefficients falls into a form of
\begin{align*}
  &\Psi^{(e)} = \begin{pmatrix} a & -b^* & c & d^*\\b & a^* & d & -c^*\\ c & d^* & a & -b^*\\d & -c^* & b & a^* \end{pmatrix}, \quad \Psi^{(o)} = \begin{pmatrix} a' & -b'^* & -c' & -d'^*\\b' & a'^* & -d' & c'^*\\ c' & d'^* & -a' & b'^*\\d' & -c'^* & -b' & -a'^* \end{pmatrix},
\end{align*}
where $\Psi^{(e)} \in \realeven$ and $\Psi^{(o)} \in \realodd$. $a,b,c,d$ and $a',b',c',d'$ are complex numbers. From above formula it is clear that we have degrees of freedom to fix one column of matrix arbitrarily in the subalgebras.

We may note, in passing, that the same is true for any fixed column, to look up the first columns, of odd base for instance,
\begin{alignat}{6}
  \gamma^0f &= \begin{pmatrix}1\\0\\0\\0\end{pmatrix},~~& \gamma^0\gamma^2\gamma^1f &= \begin{pmatrix}i\\0\\0\\0\end{pmatrix}, ~~&
  \gamma^0\gamma^3\gamma^1f &= \begin{pmatrix}0\\1\\0\\0\end{pmatrix},~~&
  \gamma^0\gamma^3\gamma^2f &= \begin{pmatrix}0\\i\\0\\0\end{pmatrix}, \nonumber\\
  -\gamma^3f &= \begin{pmatrix}0\\0\\1\\0\end{pmatrix},~~&
  -\gamma^3\gamma^2\gamma^1f &= \begin{pmatrix}0\\0\\i\\0\end{pmatrix},~~&
  -\gamma^1f &= \begin{pmatrix}0\\0\\0\\1\end{pmatrix},~~& -\gamma^2f &= \begin{pmatrix}0\\0\\0\\i\end{pmatrix}. \label{eqn:colbase}
\end{alignat}
One may see the definite role of $\gamma^2\gamma^1$ as a mediator of real and imaginary base.

It can be shown that, in fact, terms with different base, including both even and odd ones, and with coefficients in $\mathbb{C}$, i.e. in the full algebra $L^*$, are not like terms to each others. Let us first verify the feature in the complex subalgebra $\Lambda^*_-$. Since base in each line of \eqref{eqn:oddbase} does not share any non-zero position to base in other lines, it is clear that the combination with $\gamma^0$ and $\gamma^{012}$, for example, are not like terms to remaining terms. Without loss, let $\Psi = z\gamma^0 + w\gamma^{012}$ for some complex numbers $z$ and $w$. Suppose $\Psi=z'\gamma^0 + w'\gamma^{012}$ for another complex numbers $z',w'$. One easily sees that $z=z'$ and $w=w'$ since $z-iw=z'-iw'$ from the first and the third column of $\Psi$, and  $z+iw=z'+iw'$ from the second and the fourth column. Thus $z=z'$ and $w=w'$ and $\Psi$ admits unique decomposition. The same arguments apply to other base in $\Lambda^*_-$ and the ones in $\Lambda^*_+$. We should not overlook, however, that these arguments are not applied to the column Spinor with base \eqref{eqn:colbase}.

Since $z$ and $w$ here are complex numbers, $z+iw$ and $z-iw$ can be arbitrarily different complex numbers. Therefore the general form of the matrices are
\begin{align*}
  &\Psi^{(e)} = \begin{pmatrix} a_1 & a_2 & c_1 & c_2\\b_1 & b_2 & d_1 & d_2\\ c_1 & c_2 & a_1 & a_2\\d_1 & d_2 & b_1 & b_2 \end{pmatrix}, \quad \Psi^{(e)} = \begin{pmatrix} a_1' & a_2' & -c_1' & -c_2'\\b_1' & b_2' & -d_1' & -d_2'\\ c_1' & c_2' & -a_1' & -a_2'\\d_1' & d_2' & -b_1' & -b_2' \end{pmatrix},
\end{align*}
where $\Psi^{(e)} \in L^*_+$ and $\Psi^{(o)} \in L^*_-$. It is clear from above that we have degrees of freedom to fix two columns of matrix arbitrarily.

Finally we are ready to consider a $\Psi  = \Psi^{(e)} + \Psi^{(o)}\in L^*$. 
Similarly as before, assume $\Psi=\Phi^{(e)}+\Phi^{(o)}$, for some other $\Phi^{(e)}\in L^*_+$ and $\Phi^{(o)}\in L^*_-$. Let $c_i$, $i=1,2,3,4$ be the $i$-th column of $\Psi$. If $x$ denotes the first column of the $\Psi^{(e)}$ and $y$ denotes the first column of the $\Psi^{(o)}$,  $x+y=c_1$ and $\gamma^5x-\gamma^5y = c_3$ and thus $x$ and $y$ are uniquely determined. With Similar reason, the second column of $\Psi^{(e)}$ and of $\Psi^{(o)}$ are uniquely determined. Thus both matrices are uniquely determined, there is no other decomposition. Conversely, at last we see full degrees of freedom to fix a matrix arbitrarily.

Now we are able to conclude that at least with Dirac base, $L^*$ has same ability as $\Lambda^*$ to treat any complex differential forms with respect to the addition and the Clifford multiplication. One can safely state from discussions in this Section that doing algebras in $L^*$ and $\Lambda^*$ are different only in their symbols.

\subsection{Corresponding ILK equation and the modified equation}

Now we are able to investigate the corresponding ILK equation and the corresponding modified equation written on the $4\times4$ matrix Spinors.

Consider an equation
\begin{align}
  i\partial_\mu\gamma^\mu \Psi - m\Psi = 0, \label{eqn:ILK4x4}
\end{align}
where $\Psi\in L^*$ is given as in \eqref{eqn:gen4x4}. This equation has explicit $4$ Dirac equations for each column of $\Psi$. If wants, one can multiply $f$ on the right,
\begin{align*}
  i\partial_\mu\gamma^\mu \Psi f - m\Psi f = i\partial_\mu\gamma^\mu \psi - m\psi = 0
\end{align*}
and the Dirac equation is recovered for the column $\Psi f$ but now we lose the tensor-like transformation rules.
One may instead multiply, for instance, $F_1=\begin{pmatrix}1 & 0 & 0 & 0\\0 & 0 & 0 & 0\\0 & 0 & 0 & 0\\0 & 0 & 0 & 0\end{pmatrix}$, without reducing $\Psi$ to column. One may consider $F_j$, $j=1,2,3,4$, the meaning of subscript $j$ is clear. The matrix of the form, for instance $\Psi F_3$, of the form $\begin{pmatrix} 0 & 0 & a & 0 \\ 0 & 0 & b & 0 \\ 0 & 0 & c & 0 \\ 0 & 0 & d & 0 \end{pmatrix}$ constitutes one left minimal Ideal in the matrix algebra $L^*$.  To put it the other way round, the decomposition of the matrices of the form will inform us what is the left minimal Ideal in $\Lambda^*$.

Now let us consider the equation corresponding to the modified equation \eqref{eqn:dke}.
It will be
\begin{align}
  \partial_\mu\gamma^\mu\Psi - m\Psi\gamma^{012} = 0, \quad \Longleftrightarrow \quad
  \partial_\mu\gamma^\mu\Psi\,\gamma^{21} - m\Psi\gamma^0 = 0, \label{eqn:dke4x4}
\end{align}
where $\Psi$ either in $\realeven$ or in $\realodd$.
Here it is the case $\bar\theta = dz$ in \eqref{eqn:dke} without loss of generality.
At this point, we attribute the originality of the equation \eqref{eqn:dke4x4} to Hestenes \cite{hestenes_real_1967}. Hestenes wrote the same equation after a few years after the K\"ahler's work \cite{kahler_innere_1962} in his own language. His theory became known as the theory of Space-Time Algebra or Geometric Algebra, which is in fact, the Clifford algebra in Minkowski space. The work did not associated to the studies on the anti-symmetric tensor field based ILK equation. Among the superfluous mathematical languages in this subject, in fact there is only one eventuation, the Clifford algebra. Although it seems a close study was made in the paper along similar lines we are advancing here, what does not seem to be enough, however, is the explanation for the covariance of the equation and the relevance of the equation, which we shall present our own in Section \ref{sec:cov}.

Let us return to the inquiry on the equation. If we multiply $f$, one can easily find it is the Dirac equation because,
\begin{align*}
  &\gamma^{21}f = i\,f, \quad \gamma^0f = f, \quad \text{hence}\\
  &\partial_\mu\gamma^\mu\Psi\,\gamma^{21}f - m\Psi\gamma^0f = i\partial_\mu\gamma^\mu\psi - m\psi = 0.
\end{align*}
It has to be explained that the system \eqref{eqn:dke4x4} is not an over-determined one because if we restrict $\Psi$ in $\realeven$ or $\realodd$, $\Psi$ is determined by fixing the first column as noticed in the preceding Section. Thus we have $4$ degrees of freedom in complex dimensions, while the equation \eqref{eqn:dke4x4} has $4\times4=16$ complex equations. We shall show the other three columns are redundant but harmless, the equations acquired from first column are exactly reproduced from other columns in \eqref{eqn:dke4x4} so that this system is not over-determined.
If we expand the terms with the $\Psi\in\realeven$, we get
\begin{alignat}{5} \label{eqn:diracsystem}
  0=(\gamma^\mu \partial_\mu)\Psi - m\Psi\,\,\gamma^{012} =
  &\big(\partial_t \Psi_{0~~} &&+ \partial_x \Psi_{tx} &&+ \partial_y \Psi_{ty} &&+ \partial_z \Psi_{tz}+m\Psi_{xy}\big)\gamma^0\\\nonumber
  +&\big(\partial_t \Psi_{tx} &&+ \partial_x \Psi_{0~~} &&+ \partial_y \Psi_{xy} &&- \partial_z \Psi_{zx}+m\Psi_{ty}\big)\gamma^1\\\nonumber
  +&\big(\partial_t \Psi_{ty} &&- \partial_x \Psi_{xy} &&+ \partial_y \Psi_{0~~} &&+ \partial_z \Psi_{yz}-m\Psi_{tx}\big)\gamma^2\\\nonumber
  +&\big(\partial_t \Psi_{tz} &&+ \partial_x \Psi_{zx} &&- \partial_y \Psi_{yz} &&+ \partial_z \Psi_{0~}-m\Psi_{4~}\big)\gamma^3\\\nonumber
  +&\big(\partial_t \Psi_{xy} &&- \partial_x \Psi_{ty} &&+ \partial_y \Psi_{tx} &&+ \partial_z \Psi_{4~}-m\Psi_{0~}\big)\gamma^0\gamma^1\gamma^2\\\nonumber
  +&\big(\partial_t \Psi_{yz} &&+ \partial_x \Psi_{4~~} &&- \partial_y \Psi_{tz} &&+ \partial_z \Psi_{ty}-m\Psi_{zx}\big)\gamma^0\gamma^2\gamma^3\\\nonumber
  +&\big(\partial_t \Psi_{zx} &&+ \partial_x \Psi_{tz} &&+ \partial_y \Psi_{4~~} &&- \partial_z \Psi_{tx}+m\Psi_{yz}\big)\gamma^0\gamma^3\gamma^1\\\nonumber
  +&\big(\partial_t \Psi_{4~~} &&+ \partial_x \Psi_{yz} &&+ \partial_y \Psi_{zx} &&+ \partial_z \Psi_{xy}+m\Psi_{tz}\big)\gamma^1\gamma^2\gamma^3,
\end{alignat}
Since these odd base are {\it not like terms} in the matrix addition, each coefficient in \eqref{eqn:diracsystem} should be zero and $8$ independent equations are obtained. The point to observe is that considering each column separately does not have meaning in above. This $8$ equations must be the same equations the Dirac equation gives, in considering of the first column.

It is a tautology to perform a calculation in the languages of differential forms but we illustrate them.
\begin{alignat}{5} \label{eqn:dkesystem}
  0=(d-\delta)\varphi - m\varphi\vee dt\vee dx& \vee dy \nonumber\\
  =&\big(\partial_t \varphi_{0~~} &&+ \partial_x \varphi_{tx} &&+ \partial_y \varphi_{ty} &&+ \partial_z \varphi_{tz}+m\varphi_{xy}\big)dt\\\nonumber
  +&\big(\partial_t \varphi_{tx} &&+ \partial_x \varphi_{0~~} &&+ \partial_y \varphi_{xy} &&- \partial_z \varphi_{zx}+m\varphi_{ty}\big)dx\\\nonumber
  +&\big(\partial_t \varphi_{ty} &&- \partial_x \varphi_{xy} &&+ \partial_y \varphi_{0~~} &&+ \partial_z \varphi_{yz}-m\varphi_{tx}\big)dy\\\nonumber
  +&\big(\partial_t \varphi_{tz} &&+ \partial_x \varphi_{zx} &&- \partial_y \varphi_{yz} &&+ \partial_z \varphi_{0~}-m\varphi_{4~}\big)dz\\\nonumber
  +&\big(\partial_t \varphi_{xy} &&- \partial_x \varphi_{ty} &&+ \partial_y \varphi_{tx} &&+ \partial_z \varphi_{4~}-m\varphi_{0~}\big)dt\vee dx\vee dy\\\nonumber
  +&\big(\partial_t \varphi_{yz} &&+ \partial_x \varphi_{4~~} &&- \partial_y \varphi_{tz} &&+ \partial_z \varphi_{ty}-m\varphi_{zx}\big)dt\vee dy\vee dz\\\nonumber
  +&\big(\partial_t \varphi_{zx} &&+ \partial_x \varphi_{tz} &&+ \partial_y \varphi_{4~~} &&- \partial_z \varphi_{tx}+m\varphi_{yz}\big)dt\vee dz \vee dx\\\nonumber
  +&\big(\partial_t \varphi_{4~~} &&+ \partial_x \varphi_{yz} &&+ \partial_y \varphi_{zx} &&+ \partial_z \varphi_{xy}+m\varphi_{tz}\big)dx\vee dy \vee dz.
\end{alignat}

In conclusion, the equation \eqref{eqn:dke4x4} does not carry any more information than the column-wise equation
\begin{align*}
  \partial_\mu\gamma^\mu\Psi\,\gamma^{21}F_j - m\Psi\gamma^0F_j = 0,
\end{align*}
and each is the same system of $8$ partial differential equations attained from \eqref{eqn:diracsystem} and from the Dirac equation.

\subsection{Geometric lifting of Dirac Spinor}
Let us now return to the Dirac equation and the Dirac Spinor. The Dirac Spinor has degrees of freedom to put one complex column vector arbitrarily. We adopted $4\times4$ matrix Spinor because of their tensor interpretation but we are in stage to reduce our degrees of freedom as much as the Dirac Spinor. There must be a $4\times4$ matrix Spinor for a given Dirac Spinor, once one accepts the Dirac Spinor could attain its geometric primitive. The matrix should be distinct for each Dirac Spinor as seen from the sound role of the Dirac Spinor in particle physics. The problem is, however, there are too many; infinitely many $4\times4$ matrix Spinors have a same first column, and further the four columns do not seem to have distinguished role. Thus we are faced to the problem to determine a consistent way to single out one matrix Spinor from one Dirac Spinor. This amounts to assigning unique geometric sense on each Dirac Spinor, for the procedure will determine the unique element in $\Lambda^*$ also.

As has been touched in the preceding Section, we shall examine two possibilities. One is to use matrices of the form
\begin{align*}
  \Psi = \begin{pmatrix} a & 0 & 0 & 0\\b & 0 & 0 & 0\\ c & 0 & 0 & 0\\d & 0 & 0 & 0 \end{pmatrix}.
\end{align*}
The matrices of above form constitute a left minimal Ideal in the $L^*$. What has to be noticed is that one cannot generate such a matrix without exploiting the full algebra $L^*$. To have such a matrix, it calls for both of aspects, all the even and odd base are to be used, and the coefficients must be complex numbers. This is clear from the discussion in previous Section. Thus projection step is necessarily followed among $L^*$.

The other possibility is to use matrices of the form
\begin{align*}
  \Psi = \begin{pmatrix} a & -b^* & c & d^*\\b & a^* & d & -c^*\\ c & d^* & a & -b^*\\d & -c^* & b & a^* \end{pmatrix}.
\end{align*}
The relevance of the way filling other three columns comes from the fact that above matrices constitute a subalgebra $\Realeven$. One may equivalently consider the $\Realodd$ by considering
\begin{align*}
  \Psi = \begin{pmatrix} a' & -b'^* & -c' & -d'^*\\b' & a'^* & -d' & c'^*\\ c' & d'^* & -a' & b'^*\\d' & -c'^* & -b' & -a'^* \end{pmatrix}.
\end{align*}

While the first approach directly gives the equivalence to the Dirac equation as seen in the preceding Section, the second approach is not. Recall our earlier examination that an element of the even subalgebras $\Lambda^*_+$ cannot be a solution of the ILK equation. The same is true for this corresponding equation.

However one also noticed that the massless Dirac equation fits to this approach too well in the Section \ref{sec:even}, in both continuum and discrete geometries. It is also noteworthy as discussed by Benn and Tucker \cite{benn_dirac_1985} that the subalgebra separation is safely achievable in the general space-time.

Besides of these, there is one more aspect. While the Ideal separation is so easily computed in the $4\times4$ matrix formalism, the column separation, the separation in the $\Lambda^*$ is not as trivial as those or as the subalgebra separation. The projected field is a combination of all base over complex numbers and it seems there is not much helpful symmetry. It would be better to say that the geometric meaning assinged by lifting a Dirac Spinor to a matrix one in an Ideal seems to be biased with the pure mathematics adapted to the matrix formalism not with the geometric considerations. We do not know, however, what geometric features the half-spin particles should have and we must say our approach also lies in purely mathematical grounds. Nevertheless, this non-trivial separation is a rather conceivable feature in the Lattice theory.

With these points, it is worth to investigate the possibility to keep the separation procedure simple with the subalgebra separation but to modify the ILK equation. Even if the geometric lifting procedure is purely mathematical, one might use any choice preferred.

\section{Covariance of the Modified Equation} \label{sec:cov}
We are now in stage to demonstrate the most crucial issue, the modification is not any more {\it ad-hoc} than the original Dirac equation and is one achievable formalism for the half-spin particles with observables of momentum, position and spins.

First of all, it is not hard to explain why there are degrees of freedom to choose $\bar\theta$. It is because the original Dirac equation has degrees of freedom to choose Gamma matrices. Let us consider new Gamma matrices given by relations
\begin{align*}
  \tilde{\gamma}^\mu:=S^{-1} \gamma^\mu S, \quad \mu=0,1,2,3
\end{align*}
for some $S$. For $S = S[\Lmn]$, $\tilde{\gamma}^\mu = \Lmn \gamma^\nu$. Let $\bar\theta = dz$. If one fixes $\bar\theta'_\mu = (\Lambda^{-1})^\nu_{\,\,\,\,\mu}(\bar\theta)_\nu=(\Lambda^{-1})^\nu_{\,\,\,\,\mu}(dz)_\nu$, and $S=S[\Lmn]$, then in the equation \eqref{eqn:dke4x4}, we replace $\gamma^{012}$ attached by $S\gamma^{012}S^{-1}$. The result is
\begin{align*}
  &\partial_\mu\gamma^\mu\Psi - m\Psi\,\,\big(S\gamma^{012}S^{-1}\big) = 0 \\
  \Longleftrightarrow\quad &S^{-1}\big(\partial_\mu\gamma^\mu\Psi - m\Psi S^{-1}\gamma^{012}S\big)S=0, \\
  \Longleftrightarrow\quad &\partial_\mu (S^{-1}\gamma^\mu S) \,\, (S^{-1}\Psi S) - m (S^{-1}\Psi S)\,\,\gamma^{012}=0. 
\end{align*}
Thus the equation now is the equation of an original form \eqref{eqn:dke4x4} with different Gamma matrices and $\Psi$ defines in terms of new Gamma matrices but with same coefficient matrix $\gamma^{012}$. $\gamma^{012}$ has different meaning now, however, it does not correspond to $dt\vee dx \vee dy$ anymore but to $*(\bar\theta')$. Since we are free to choose Gamma matrices, $\bar\theta$ must be free parameter.

Then there arises a question that what is the meaning of the fixing one direction $\bar\theta$. For some observer $\bar\theta$ looks complicated than the observer who sees $\bar\theta=dz$. The answer to this question comes from the form of eigenstates of spin observables. In the Dirac equation, $\gamma^3 = \begin{pmatrix} 0 & \sigma^3 \\ -\sigma^3 &0 \end{pmatrix}$ is related to the Spin operator to that direction $S_z$ and $S_z$ has $4$ eigenstates
\begin{align*}
  \begin{pmatrix}1 \\ 0 \\ 0 \\ 0\end{pmatrix}, \quad
  \begin{pmatrix}0 \\ 1 \\ 0 \\ 0\end{pmatrix}, \quad
  \begin{pmatrix}0 \\ 0 \\ 1 \\ 0\end{pmatrix}, \quad
  \begin{pmatrix}0 \\ 0 \\ 0 \\ 1\end{pmatrix}.
\end{align*}
The above are the standard base but are not distinguished base. Since $\gamma^3$ can be assigned to any direction considering the freedom to choose Gamma matrices, it would be better to denote in this context the $S_3$ operator. The direction $\bar\theta$ is special by the amount the $\gamma^3$ is special that the eigenstates of Spin observable to $\bar\theta$ direction retain the simple form, and nothing more. In this way, the equation is covariant and does not break the principle of relativity.

We shift the emphasis away from the free parameter $\bar\theta$ to a deeper question whether the modified equation indeed describes the half spin particles. In the following section, although the discussion will not prove all the issues, we try to answer the question.
\subsection{Eigenvalue problems in the restricted subspaces}
The underlying philosophy in this section about the Dirac equation is that it is the eigenvalue problem of the operator $\slashed{p} = p_\mu\gamma^\mu$. This $4\times4$ matrix attains pure real eigenvalues $m$ and $-m$, where $m = \sqrt{(p^0)^2 - |\mathbf{p}|^2}$ and each of them is associated to the two independent eigenstates. They are the spin up/down particles and the spin up/down anti-particles.

Now let us extend the discussion on the eigenvalue problems to the general operators. For the first, consider the simplest case, one-dimensional first order differential operator $\frac{d}{dx}$ on a real-valued function $u$. The equation for the eigenvalue problem is $u_x = \lambda u$. Then the solution will be $u = c_0\exp(\lambda x)$. One might consider $i\psi_x = \lambda \psi$ for a complex scalar field $\psi$. The solution is $\psi = c_0\exp{(-i\lambda x)}$.

Before we make any conclusion from those, let us consider one more example, the first order operator in $2$-dimensions. If we split the Cauchy-Riemann equations to the real part equation and the imaginary part equation, there arises the problem to determine the sign of the equations. For example, the imaginary part equation $-u_y + v_x = 0$ can be a $u_y - v_x = 0$. This is not the relevant question in the original equation but in its eigenvalue problem, the sign becomes important since it determines whether to add the $\lambda\begin{pmatrix} u \\ v \end{pmatrix}$ or $\lambda\begin{pmatrix} u \\ -v \end{pmatrix}$. For now, let us follow signs as
\begin{align}
  &\partial_x \begin{pmatrix} 1 & 0 \\0 & -1 \end{pmatrix} \begin{pmatrix} u \\ v \end{pmatrix} + \partial_y \begin{pmatrix} 0 & 1 \\1 & 0 \end{pmatrix} \begin{pmatrix} u \\ v \end{pmatrix} - \lambda \begin{pmatrix} u \\ v \end{pmatrix} \label{eqn:2x2}\\
  =:\,\,&\partial_x \,\,\alpha \begin{pmatrix} u \\ v \end{pmatrix} + \partial_y \,\,\beta \begin{pmatrix} u \\ v \end{pmatrix} - \lambda \begin{pmatrix} u \\ v \end{pmatrix}  = 0. \nonumber
\end{align}
One notices $\alpha = \sigma^3$ and $\beta = \sigma^1$, the Pauli matrices and hence $\alpha^2 = \beta^2 = \textrm{I}$ and $\alpha\beta + \beta\alpha = 0$. The reason for choosing signs in \eqref{eqn:2x2} is its eigenvalues, which we will inqure soon.

Now we make a conclusion with above examples. While the second order operator $(d-\delta)^2 = \Delta$ translates to the languages of differential geometry without any elaborations the first order operator $(d-\delta)$ does not. Suppose one wants to confine the space the operators acts on to only scalars. Since $d$ increases the order of differential form by one and $\delta$ decreases the order by one, $(d-\delta)^2 = -(\delta d+d\delta)$ does not alter the order and hence its eigenvalue problem is well-defined, $(d-\delta)^2 u - \lambda u=0$.

For the first order operators, however, it seems one cannot consider an eigenvalue problem in a certain restricted subspace. In $(d-\delta)u - \lambda u =0$, since the order of two terms are different the equation is vacuous. Bearing in mind that there is no reason not to consider an eigenvalue problem of first order operators, especially with a certain degrees of freedom, for example within real scalar fields, this must be quite weird. The resolution is rather easy, if the equation $u_x - \lambda u=0$ tells us some truth, then it means $(d-\delta)u -  \lambda u \vee dx =0$ does same. If wants, one can instead regards this as the eigenvalue problem of the operator
\begin{align}
  \varphi \mapsto (d-\delta)\varphi\vee dx. \label{eqn:1storder}
\end{align}
One might claim this an {\it ad-hoc} arguments, but we merely say that the \eqref{eqn:1storder} is a well-defined operator on real scalars.

We find the $2$-dimensional examples similar. If we connect the vector field $\begin{pmatrix} u\\ v\end{pmatrix}$ to $u\,dx+ v\,dy$, after differentiations in the equation \eqref{eqn:2x2}, one sees in the expression
\begin{align*}
  \begin{pmatrix} u_x \\ -v_x \end{pmatrix} + \begin{pmatrix} v_y \\u_y \end{pmatrix}
\end{align*}
that the first component of the vector corresponds to $0$-form and the second component to density $dy\wedge dx$; $u_x \, dx \vee dx = u_x$, $v_y \, dy \vee dy =v_y$, $v_x \, dx \vee dy = -v_x dy\wedge dx$ and $(u_y\, dy) \vee dx = u_y dy\wedge dx$. What I try to say is that if the term $\lambda\begin{pmatrix} u\\ v\end{pmatrix}$ is added in the equation then it is understood in this context that $\lambda(u\, + \,v \, dy\wedge dx) = \lambda\big(u\,dx + v\, dy\big)\vee dx$ is added. In other words, we again are dealing with an eigenvalue problem of an operator
\begin{align*}
  \varphi \mapsto (d-\delta)\varphi \vee dx
\end{align*}
acting on $1$-forms.

So far, we have inquired about the well-defined differential operators of first and second order but were not concerned with their eigenvalues. To be a consistent field theory, the operator should attain only pure real eigenvalues. For the $u_x = \lambda u$ case, it is not suitable though, the solution $c_0\exp{(\lambda\,x)}$ is not an element of $L^2$ space of course but is neither in the tempered distribution and thus does not have a well-defined Fourier transform. Therefore it is more relevant to consider $i\psi_x = \lambda \psi$ which attains only pure real eigenvalues in the space of tempered distributions, but resulting in doubling the degrees of freedom from a real $u$ to a complex $\psi$.

Nextly, the system \eqref{eqn:2x2} first seems to have only pure imaginary eigenvalues since for the $\begin{pmatrix} u\\v\end{pmatrix} = \begin{pmatrix} u_0\\v_0\end{pmatrix} e^{-i\mathbf{p}\cdot \mathbf{x}}$,
\begin{align*}
  0=\textrm{det}\,\, \begin{pmatrix} -ip^x -\lambda & -ip^y \\ -ip^y & ip^x-\lambda\end{pmatrix} = \lambda^2 + |\mathbf{p}|^2.
\end{align*}
In these considerations, it would be relevant to consider
\begin{align*}
    i\left(\partial_x \alpha  + \partial_y \beta \right) \begin{pmatrix} \psi_1 \\ \psi_2 \end{pmatrix} - \lambda \begin{pmatrix} \psi_1 \\ \psi_2 \end{pmatrix} = 0
\end{align*}
with $i$ attached. There, we again doubled the degrees of freedom because $\psi_1$ and $\psi_2$ are complex-valued now. The $1+1$-dimensional Dirac equation is formulated in this way, with this amounts of degrees of freedom.

The arguments saying \eqref{eqn:2x2} has pure imaginary eigenvalues are mistake, however, we can manage the degrees of freedom not to be increased. This is an easy consequence of tieing $u$ and $v$ to form a  $\psi = u+iv$. If $\psi = (u_0+iv_0)e^{-is}$, then $\begin{pmatrix} u \\ v \end{pmatrix} = \begin{pmatrix} \cos s & \sin s \\ -\sin s & \cos s \end{pmatrix} \begin{pmatrix} u_0 \\ v_0 \end{pmatrix}$. Here $s=\mathbf{p}\cdot \mathbf{x}$. The calculation shows this is an eigenstate with real eigenvalue $\lambda$ such that $\lambda^2 = |\mathbf{p}|^2$.

We inquire further on these eigenstates. The rotation matrix
\begin{align*}
  &\begin{pmatrix} \cos s & \sin s \\ -\sin s & \cos s \end{pmatrix} \\
  &= \exp{(s\alpha\beta)}= \big(1 - \frac{s^2}{2!} + \frac{s^4}{4!} - \cdots\big)\textrm{I} + \big(s - \frac{s^3}{3!} + \frac{s^5}{5!} - \cdots\big)\alpha\beta\\
  &= \cos s\,\, \textrm{I} + \sin s\,\, \alpha\beta.
\end{align*}
Now let $G$ be a group of operators from $\mathbb{R}^2$ to $\mathbb{R}^2$. Then $\alpha\beta$ is an element of the Lie algebra of the group $G$. Note that $i$ does not belong to the Lie algebra of $G$ but is in the Lie algebra of groups $\mathbb{C}^2 \mapsto \mathbb{C}^2$ or $\mathbb{C} \mapsto \mathbb{C}$. In other words, the use of $\exp{(-is)}$ implicitly enlarges the space of states. Since we here want to manage the degrees of freedom precisely, it should not be employed.

Furthermore, since $\alpha\beta$ is the unique non-trivial element of the Lie algebra of the subgroup $O(2)$, $\exp{(s\alpha\beta)}\begin{pmatrix} u_0 \\ v_0 \end{pmatrix}$ stays in the reasonable space of states and works well as an eigenstate of the operator.
\subsection{The modified equation : a well-defined eigenvalue problem on differential forms}
Now we are able to demonstrate our point of view on the modified equation. To start with, consider $p_\mu\, dx^\mu$ corresponding to $\slashed{p}$. We are supposed to work within the restricted subalgebras. As demonstrated in the preceding Section, for this first order operator in the restricted subalgebra, the eigenvalue problem of $p_\mu\, dx^\mu$ can be written down, once we determine which component of differential form $\varphi$ corresponds to a certain position in the $\mathbb{C}^4$ vector $\psi$,

To be concrete, if we match up a certain Gamma matrices $\gamma^\mu$ to $dx^\mu$, and presume that $\psi = \Psi\,f$ and $\Psi$ is given by the formula \eqref{eqn:gen4x4}, then we have all necessary information and are able to write down the equation of the eigenvalue problem. If Dirac base is used, for instance, the translated equation for the eigenvalue problem in that case is
\begin{align}
  p_\mu dx^\mu \varphi - \lambda\varphi \vee dt=0.
\end{align}
%

Next, we examine the corresponding operator of $\slashed{\partial}=\partial_\mu \gamma^\mu$ and its eigenstates. When we investigate the eigenvalues of $\slashed{\partial}$, $\psi_0 e^{-ip\cdot x}$ is tested for the eigenvector. To be radical, this can be justified since $i$ is in the Lie algebra of the group of operators $\mathbb{C}^4 \mapsto \mathbb{C}^4$ so that $\exp{(-ip\cdot x)}\psi_0$ is a legal expression.

In inspection of eigenstates of $(d-\delta) = \partial_\mu dx^\mu$ on $\realeven$ or $\realodd$, to work strictly within those subalgebras, it is clear that $i$ should be abandoned to exponentiate for the similar reason as before. Instead, let us use $\chi$. We are not fixed what is $\chi$ yet. For simple discussion, let us confine $\chi$ in a limited form, a spatial coordinate basis $2$-form with unit norm, such as $dx \vee dy, dy \vee dz, \cdots$. Also $\chi'$ denotes the inverse of $\chi$. We then try
\begin{align*}
  \varphi(\mathbf{p}) = \bar\varphi\exp{\big(-s\chi\big)} = \bar\varphi\exp{\big(s\chi'\big)}=\varphi\cos s \,\,\textrm{I} + \sin s \,\,\chi', \quad s=p\cdot x.
\end{align*}
as an eigenstate. The meaning of $4$-vector $p$ is clear.

In comparison to earlier two dimensional case, it is clear that $\chi$ tells us which components of $\varphi$ are tied together to form a complex number. That is to say we are following the reverse logic. Let us denote $0$-form, $2$-forms and $4$-form component of $\varphi$ by $\varphi_0$, $\varphi_{\mu\nu}$ and $\varphi_{txyz}$. Suppose $\chi=dy\vee dz$ for instance, then it means that $\varphi_0$ component is tied with $\varphi_{yz}$ to form a complex number, and that $\varphi_{tx}$ is tied with $\varphi_{txyz}$ and so on.

We claim that this tieing process, results from fixing $\chi$, provides us a well-defined procedure to choose one set of Gamma matrices, in particular, among the Dirac base. In inspection on the Dirac base in Section \ref{sec:spinor}, there is only one basis which has the same role of $\chi$, as a mediator of real and imaginary base, the $\gamma^2\gamma^1$. Thus we are supposed to set $\gamma^2$ corresponds to $dy$, the first part of $\chi$ and $\gamma^1$ to $dz$, the second part of $\chi$. It is certain that $\gamma^0$ should be assigned for $dt$ according to the sign of metric and the last $\gamma^3$ has to be assigned for the remaining $dx$. This should be regarded as a convenient heuristic, however, since we chose Gamma matrices among the Dirac base only and even only as a single one. Nevertheless, above procedure is well-defined in our limited situation.


Lastly, let us put together above observations. Calculation shows
\begin{align*}
  (d-\delta)\varphi(\mathbf{p})
  &= (d-\delta)\varphi_0\big(\cos s\,\textrm{I} + \sin s \chi') \\
  &= dx^\mu \vee \partial_\mu \, \varphi_0\big(\cos s\, \textrm{I} + \sin s \chi'\big) \\
  &= p_\mu \, dx^\mu \vee \varphi_0\big(-\sin s\, \textrm{I} + \cos s \chi'\big), \quad\text{where $s=p\cdot x$} \\
  &= p_\mu \, dx^\mu \vee \varphi_0\big(\cos s\, \textrm{I} + \sin s \chi'\big)\chi' \\
  &= p_\mu \, dx^\mu \vee \varphi(\mathbf{p}) \vee \chi'.
\end{align*}
Therefore
\begin{align}
  (d-\delta)\varphi(\mathbf{p})\vee\chi = p_\mu \, dx^\mu \vee \varphi(\mathbf{p}). \label{eqn:eigenphi}
\end{align}
Note the right-hand-side of \eqref{eqn:eigenphi} is the operator $p_\mu \, dx^\mu$ we matched with $\slashed{p}$. Since we have determined Gamma matrices, we can think of the eigenvalue problem of it corresponding to one of $p_\mu\gamma^\mu$. The operator has a well-defined equation for the eigenvalue problem,
\begin{align*}
  (d-\delta)\varphi \vee\chi - m\varphi \vee dt=0
\end{align*}
and has two pure real eigenvalues.

The most important point to observe is that from the eigenstate $\varphi_0\exp{\big(-(p\cdot x)~ \chi\big)}$, we retrieve the momentum by $\partial_\mu \,\,\big( \varphi_0\exp{\big(-(p\cdot x)~\chi\big)}\,\,\big) \vee \chi$. At this stage, we arrive at the genuine conclusion: the modified theory develops a Quantum Mechanics of half-spin particles with the uncertainty principle implemented with the commutation relation
\begin{align}
  \big[\mathbf{P},\,\mathbf{Q}] = \chi \label{eqn:uncertainty}
\end{align}
and momentum operator
\begin{align}
  \mathbf{P}_\mu : \varphi \mapsto \partial_\mu \varphi \vee \chi \label{eqn:Poperator}
\end{align}
with eigenstates of a form
$$\varphi_0\exp{\big(-(p\cdot x)~ \chi\big)}$$
in $\realeven$ or $\realodd$. In other words, by the states $|\mathbf{p},\uparrow\rangle,|\mathbf{p},\downarrow\rangle, \cdots$, we mean the
\begin{align}
  \varphi_0^{\mathbf{p},\uparrow}e^{-(p\cdot x)~ \chi},\quad\varphi_0^{\mathbf{p},\downarrow}e^{-(p\cdot x)~ \chi},\quad \cdots, \label{eqn:eigstates}
\end{align}
instead of
$$\psi_0^{\mathbf{p},\uparrow}e^{-i(p\cdot x)},\quad\psi_0^{\mathbf{p},\uparrow}e^{-i(p\cdot x)},\quad \cdots.$$
In fact, these and the first columns of matrices in \eqref{eqn:eigstates} coincides and the latter does not have any more degrees of freedom.

We shall not look up all the commutation and anti-commutation relations of all observables of the half-spin particles and not build a Quantum Mechanics here. It is beyond the scope of this paper.

\subsection{Summary}
We summarize the conclusions of this Section and remark a few. First and foremost, it has to be addressed that although this paper has been devoted to inquire a different mathematical object from Dirac Spinor, little are changed in the use. As a consequence of what have seen so far, we do not need to keep the modified equation instead of the Dirac equation in every concerns of the particle with only one exception. To take an example, there are no other plane wave solutions of the modified equation other than the corresponding ones the Dirac equation gives us since the first column of matrix state fixes the remaining columns. This is an important difference from the original ILK theory. The one exception is the transformation rule under the Lorentz group. What has to be done for the modified theory when we are in position to take a transformation of the Dirac equation is, to recall the equation was the concise compaction of the modified equation and to revoke other three columns and also to regard $i$ as the first column of $\gamma^2\gamma^1$ multiplied from right. In other words, we take one scenario of geometric lifting of the equation before taking a transformation. After the transformation, we may again discard the three columns.

It would be phrased in the different angle. The three equations discussed in this paper, the Dirac equation, the modified equation and the modified equation in $4\times4$ matrix Spinors are all expanded to the same $8$ linearly independent equations in \eqref{eqn:diracsystem} in a fixed coordinate system although they look different. In other words, they are the identical partial differential equations in that coordinate system. They, however, transform as their own rule.

Nevertheless, the local observables of the fields will remain same. Since equations were same, the local covariant observables made out from the solution of each equation must be same. Since the local observables are tensors, not Spinors, they transform as tensors and hence every observer will see the same consistent observables.

Now we remark two things. First, we avoided to use the projection operators onto scalar, vector, and higher order components of $\varphi \in \Lambda^*$  and also the involution operators defined on $\Lambda^*$. What used so far is the Hodge-star operator $*$ and the two products $\wedge$ and $\vee$. Secondly, although we only touched on the subject in Section \ref{sec:cov}, the modified equation seems to be settled by the first principle approach, with commutation relations. The modified equation has no other meaning.
\section{Chiral Theory of the Modified Equation} \label{sec:chiral}
Last Section discusses how the chirality, the another important feature of the Dirac theory, is realized in the language of modified equation.

In the Dirac theory, the chirality is expressed with aid of the fifth Gamma matrix $\gamma^5$, defined by
\begin{align*}
  \gamma^5 = i\gamma^0\gamma^1\gamma^2\gamma^3 = i\frac{1}{4!}\epsilon_{\mu\nu\alpha\beta}\gamma^\mu\gamma^\nu\gamma^\alpha\gamma^\beta.
\end{align*}
$\gamma^5$ has the following properties,
\begin{align}
  \big(\gamma^5\big)^2 = \textrm{I}, \quad \{\gamma^5,\,\,\gamma^\mu\} = 0, \quad\text{for $\mu=0,1,2,3$}. \label{eqn:G5prop}
\end{align}
Now one can project a state into its left chiral part and the right chiral part, the anti-self-dual and the self-dual part with respect to the action of $\gamma^5$
\begin{align*}
  &\psi = \psi_L + \psi_R, \\
  &\psi_L = \frac{\psi - \gamma^5\psi}{2}, \quad \psi_R = \frac{\psi + \gamma^5\psi}{2}.
\end{align*}
If $\psi$ is a solution of the massless Dirac equation then so do both of parts. Furthermore, their phase shifted ones $e^{i\theta_L}\psi_L$ and $e^{i\theta_R}\psi_R$, with $\theta_L\ne\theta_R$ also are solutions of the massless Dirac equation. These constitute the Axial U(1) symmetry. Also the Dirac equation splits,
\begin{align*}
  &i\partial_\mu \gamma^\mu \psi_L - m\psi_R = 0 \\
  &i\partial_\mu \gamma^\mu \psi_R - m\psi_L = 0.
\end{align*}
It should be noticed from above that the operator $i\partial_\mu\gamma^\mu$ flips the chirality.

In the abstract level of investigation on the chirality, one realizes that two properties of \eqref{eqn:G5prop} indeed do an important role. To have a consistent chiral theory, first one should have an operator squares to identity like $\gamma^5$ and secondly one should prove that the chirality flips after operating the differential operator $i\partial_\mu\gamma^\mu$. The sign of $\big(\gamma^5\big)^2$ is crucial, if $\big(\gamma^5\big)^2$ were $-\textrm{I}$, then it would not make sense
because for $\psi' = -\gamma^5 \psi$, $\psi_L = \frac{\psi - \gamma^5 \psi}{2} = \frac{\psi' + \gamma^5 \psi'}{2}=\psi'_R$. In other words, the space of left chiral states and the space of right chiral states would be mixed up. Anti-commutativity of $\gamma^5$ with all $\gamma^\mu$ for $\mu=0,1,2,3$ enables us to prove that $i\partial_\mu\gamma^\mu$ flips the chirality.

We are now in a position to discuss chiral theory in the modified equation. As physicists have been used the terminology such as a pseudo-scalar and a pseudo-vector for quantities made up with $\gamma^5$ operated ones, for instance $\bar\psi\gamma^5 \psi$ for a pseudo-scalar, $\gamma^5$ operator does similar role as the Hodge-star operator $*$.  To be precise, they are slightly differ and this was already pointed out by Rabin \cite{rabin_homology_1982}. For clarity, let us use $L^*$ and $\Psi$ defined by \eqref{eqn:gen4x4} for the discussion. It is clear one can define $*$ on $L^*$. If $\gamma^{0123} = \gamma^0\gamma^1\gamma^2\gamma^3 = -i\gamma^5$,
\begin{align*}
  \gamma^{0123} =
    \begin{cases}
      * & \text{on $0,1,4$ forms,} \\
      -* & \text{on $2,3$ forms.}
    \end{cases}
\end{align*}
This results in the following differences
\begin{align*}
  &\big(\gamma^{0123}\big)^2 = -\textrm{I}, \quad\big(\gamma^5\big)^2=\textrm{I} \quad \text{on both of even and odd subalgebras,}\\
  &(*)^2 = \begin{cases} \textrm{I} & \text{on odd subalgebra,} \\ -\textrm{I} & \text{on even subalgebra.} \end{cases}
\end{align*}
We address, in passing, that in Euclidean metric, one defines $\gamma^5 = \gamma^{1234}$ with Euclidean Gamma matrices and $\gamma^5$ coincides with the Hodge-star operator.

Rabin \cite{rabin_homology_1982} did not explain the cause of such sign differences, but its mathematical reason is clear. This is due to the following sign differences between two scalar products
\begin{align*}
  (\omega,\omega) =
    \begin{cases}
     \omega \vee \omega & \text{on $0,1,4$ forms,} \\
     -\omega \vee \omega & \text{on $2,3$ forms.}
     \end{cases}
\end{align*}
See the definition of the Hodge-star operator \eqref{eqn:*} for the use of paring in the definition.
To take examples, for $1$-form $dx$, $(dx,dx) = dx \vee dx = -1$, but for $2$-form $dx\vee dy$, $(dx\vee dy, \,dx\vee dy) = \textrm{det}\, \begin{pmatrix} -1 & 0 \\ 0 & -1 \end{pmatrix} = 1$, while $ dx\vee dy \vee dx\vee dy = -1$.

Since the each of the scalar products does not seem to be more canonical than each other, one may as well define $*'$ operator similarly to $*$ by
\begin{align}
  &*' : \Lambda^p(M^4) \mapsto \Lambda^{4-p}(M^4),  \nonumber\\
  &*'\omega \vee \omega = (dt\vee dx\vee dy\vee dz) \vee (\omega \vee \omega). \label{eqn:*prime}
\end{align}
Compare with the formula \eqref{eqn:*}. We presented the formula with same element $\omega$ to avoid a complicated definition for general cases. From above one can write $*'\,\omega = dt\vee dx\vee dy\vee dz \vee \omega$. In the algebra $L^*$, $*'\,\,\Psi$ will be $\gamma^{0123}\Psi = -i\gamma^5\Psi$.

Now we obtained the operator $*'$ doing same role as $\gamma^{0123}$ which squares to $-\textrm{I}$. We are in stage to come up with an operator doing same role as $\gamma^5 = i\gamma^{0123}$. Since we are suppressing the use of $i$ and $\chi$ was multiplied on the right instead in earlier discussions, we do that again. One may define another dual operator and the pseudo-norm
\begin{align}
  &*_5 : \Lambda^p(M^4) \mapsto \Lambda^{4-p}(M^4),  \nonumber\\
  &*_5 \omega = *'\,\,(\omega\vee \chi), \\
  &(\omega,\omega)_5 = *_5\omega \vee \omega.
\end{align}
In the $L^*$ it will be defined as
\begin{align*}
  *_5 \,\, \Psi = \gamma^0\gamma^1\gamma^2\gamma^3\,\Psi\,\gamma^2\gamma^1.
\end{align*}

We can verify
\begin{align*}
  (*_5)^2 = \textrm{I}, \quad *_5 (\gamma^\mu\Psi)  + \gamma^\mu(*_5\Psi) = 0, \quad \text{for $\mu=0,1,2,3$}
\end{align*}
and hence we are able to build a chiral theory with,
\begin{align*}
  \Psi_L = \frac{\Psi - *_5\,\Psi}{2}, \quad \Psi_R = \frac{\Psi + *_5\,\Psi}{2},
\end{align*}
the anti-self-dual and the self-dual part of $\Psi$ with respect to the $*_5$ operator.
Note again that the first columns of above matrices coincide with the ones from Dirac Spinors.

Now one can indeed prove that the differential operator flips the chirality,
\begin{align*}
  2\partial_\mu\gamma^\mu\Psi_L \gamma^2\gamma^1
  &=\partial_\mu\gamma^\mu\Psi \gamma^2\gamma^1 - \partial_\mu\gamma^\mu *_5\,\Psi \gamma^2\gamma^1 \\
  &=\partial_\mu\gamma^\mu (*_5\,*_5)\,\Psi \gamma^2\gamma^1 - \partial_\mu\gamma^\mu *_5\,\Psi \gamma^2\gamma^1 \\
  &=-*_5\,\partial_\mu\gamma^\mu *_5\,\Psi \gamma^2\gamma^1 - \partial_\mu\gamma^\mu *_5\,\Psi \gamma^2\gamma^1 \\
  &=-(a + *_5\,a), \quad \text{where $a=\partial_\mu\gamma^\mu *_5\,\Psi \gamma^2\gamma^1$}.
\end{align*}
\bibliographystyle{unsrt}
\bibliography{dke}
\end{document}